\documentclass[acmsmall]{acmart}
\usepackage{amsmath}
\usepackage{multirow}
\usepackage{lscape}
\usepackage{courier}
\AtBeginDocument{%
  \providecommand\BibTeX{{%
    \normalfont B\kern-0.5em{\scshape i\kern-0.25em b}\kern-0.8em\TeX}}}

\setcopyright{acmcopyright}
\copyrightyear{2023}
\acmYear{2023}
\acmDOI{XXXXXXX.XXXXXXX}

\acmPrice{15.00}
\acmISBN{978-1-4503-XXXX-X/18/06}

\begin{document}

%%
%% The "title" command has an optional parameter,
%% allowing the author to define a "short title" to be used in page headers.
\title{High Performance GPU Accelerated MuST Software}

%%
%% The "author" command and its associated commands are used to define
%% the authors and their affiliations.
%% Of note is the shared affiliation of the first two authors, and the
%% "authornote" and "authornotemark" commands
%% used to denote shared contribution to the research.
\author{Xiao Liang}
\email{liangstein@psc.edu}
\author{Edward Hanna}
\email{ehanna@psc.edu}
\author{Derek Simmel}
\email{dsimmel@psc.edu}
\affiliation{%
  \institution{Pittsburgh Supercomputing Center}
  \streetaddress{300 S. Craig St.}
  \city{Pittsburgh}
  \state{PA}
  \country{USA}
  \postcode{15213}
}

\author{Hang Liu}
\affiliation{%
  \institution{Texas Advanced Computing Center}
  \streetaddress{10100 Burnet Rd}
  \city{Austin}
  \state{TX}
  \country{USA}
  \postcode{78758}}
\email{hliu@tacc.utexas.edu}

\author{Yang Wang}
\authornotemark[1]
\affiliation{%
  \institution{Pittsburgh Supercomputing Center}
  \streetaddress{300 S. Craig St.}
  \city{Pittsburgh}
  \state{PA}
  \country{USA}
  \postcode{15213}
}
\email{ywg@psc.edu}
%%
%% By default, the full list of authors will be used in the page
%% headers. Often, this list is too long, and will overlap
%% other information printed in the page headers. This command allows
%% the author to define a more concise list
%% of authors' names for this purpose.
\renewcommand{\shortauthors}{Liang and Hanna, et al.}

%%
%% The abstract is a short summary of the work to be presented in the
%% article.
\begin{abstract}
  The MuST package is a computational software designed for ab initio electronic structure calculations for
solids. The Locally Self-consistent Multiple Scattering (LSMS) method implemented in MuST allows to
perform the electronic structure calculation for systems with a large number of atoms per unit cell. For the
LSMS method with muffin-tin potential approximation, the major computational challenge is the matrix inverse for the scattering matrix calculation, which could take more than 90\% of the computing time. However, the matrix inverse can be
significantly accelerated by modern graphical-processing-units (GPUs). In this paper, we discuss
our approach to the code acceleration by offloading the matrix inverse tasks to the GPUs through a
Fortran-C interface from the Fortran code to the CUDA code. We report our performance results showing significant speedup ratio achieved to the calculations of NiAu alloy, a candidate for
thermoelectric materials.
\end{abstract}

%%
%% The code below is generated by the tool at http://dl.acm.org/ccs.cfm.
%% Please copy and paste the code instead of the example below.
%%


%%
%% Keywords. The author(s) should pick words that accurately describe
%% the work being presented. Separate the keywords with commas.

%% A "teaser" image appears between the author and affiliation
%% information and the body of the document, and typically spans the
%% page.
%\begin{teaserfigure}
%  \includegraphics[width=\textwidth]{sampleteaser}
%  \caption{Seattle Mariners at Spring Training, 2010.}
%  \Description{Enjoying the baseball game from the third-base
%  seats. Ichiro Suzuki preparing to bat.}
%  \label{fig:teaser}
%\end{teaserfigure}

%\received{20 February 2007}
%\received[revised]{12 March 2009}
%\received[accepted]{5 June 2009}

%%
%% This command processes the author and affiliation and title
%% information and builds the first part of the formatted document.
\maketitle

\section{Introduction}
The Density-functional theory (DFT)\cite{Hohenberg1964B864} in the formulation of Kohn-Sham (KS) equation \cite{PhysRev.140.A1133} is widely used in studying the physical and chemical property of materials. Nowadays, directly solving the eigenvalue problem of the KS equation is widely adopted in electronic structure calculations, and it requires the approximations such as pseudopotentials (e.g., VASP, QuantumEspresso) or linearized basis sets for all-electron methods (e.g., WIEN2K) to make the size of the basis set manageable.
Different from directly solving the KS equation, the MuST package \cite{MuST} is a computational software designed for ab initio electronic structure calculations for solids based on the Multiple-Scattering theory (MST) together with the Green's function technique \cite{10.1088/2053-2563/aae7d8}.

\section{Multiple-Scattering Theory}
Solving the KS equation using MST is formulated by the Korringa-Kohn-Rostoker (KKR) method \cite{PhysRev.94.1111,KORRINGA1947392}. Different from solving the eigenvalue problem of the KS equation, in KKR method the charge density is obtained through solving the Green's function of the KS equation.

\subsection{KKR Method}
For a Schr\"{o}dinger equation, the Green's function connecting the $\textbf{r}$ and $\textbf{r}'$ is defined as:
\begin{equation}
  G(\textbf{r},\textbf{r}';\epsilon)=
  \lim_{\eta\rightarrow 0}\sum_k
  \frac{\Psi_k^*(\textbf{r})\Psi_k(\textbf{r}')}{\epsilon-\epsilon_k+i\eta}
  \label{Green_definition}
\end{equation}
where $\epsilon_k$ and $\Psi_k$ are the $k$-th eigen-energy and eigen-function, respectively.
Based on the definition, the charge density is:
\begin{equation}
  n(\textbf{r})=\frac{1}{\pi}\mathrm{Im}\int_{-\infty}^{\epsilon_F}G(\textbf{r},\textbf{r};\epsilon)d\epsilon
  \label{Density}
\end{equation}
In the framework of KKR method,
the Green's function for the $n$-th site at position $\textbf{r}_n$ is:
\begin{equation}
  G(\textbf{r}_n,\textbf{r}_n;\epsilon)=
  \sum_{L,L'}Z_L^n(\textbf{r}_{n} ;\epsilon)\tau_{L L'}^{nn}(\epsilon)
  Z_{L'}^{n*}(\textbf{r}_n;\epsilon)-
  \sum_L Z_L^n(\textbf{r}_n;\epsilon)J_L^{n*}(\textbf{r}_n;\epsilon)
  \label{Green_KKR}
\end{equation}
where $\underline{\tau}^{nn}$ is the $n$-th block of the multiple-scattering matrix $\underline{\tau}$, index $L$ is a combination of angular momentum and magnetic quantum numbers ${l,m}$,
$Z_L^n(\textbf{r}_n ;\epsilon)$ is the regular solution and $J_L^n(\textbf{r}_n ;\epsilon)$ is the
irregular solution of the Schr\"{o}dinger equation for the local effective potential at the $n$-th site. The multiple-scattering matrix $\underline{\tau}$ is obtained by the Dyson equation:
\begin{equation}
  \underline{\tau}(\epsilon)=[\underline{t}^{-1}(\epsilon)-\underline{g_0}(\epsilon)]^{-1}
  \label{Dyson_equation}
\end{equation}
with the inverse of the squared KKR matrix $\underline{M}(\epsilon)=\underline{t}^{-1}(\epsilon)-\underline{g_0}(\epsilon)$, where $\underline{t}(\epsilon)$ is the single site scattering matrix and $\underline{g_0}(\epsilon)$ is the free particle propagator between two different atom sites. When considering $N$ atoms in the unit cell and the angular momentum cutoff $l_{max}$, the rank of $\underline{M}$ are $N(l_{max}+1)^2$ without spin-canting and $2N(l_{max}+1)^2$ with spin-canting.

\subsection{LSMS Method}
The LSMS method \cite{PhysRevLett.75.2867} requires the calculation of the multiple-scattering matrix $\underline{\tau}$ for each atom in the unit cell, with the atom at the center of a cluster, called the local interaction zone (LIZ). For each atom, the Green's function is obtained in the same way as the original KKR method mentioned in Eq.\eqref{Green_KKR} and Eq.\eqref{Dyson_equation}, except that the KKR matrix in the LSMS method is built by considering the atoms in the LIZ, rather than in the entire space. As a result, in the LSMS method, the rank of the KKR matrix for each atom is $N_{\mathrm{LIZ}}(l_{max}+1)^2$ without spin-canting or $2N_{\mathrm{LIZ}}(l_{max}+1)^2$ with spin-canting, where $N_{\mathrm{LIZ}}$ is the atom number in the LIZ.

Based on the Dyson equation, the $\tau$-matrix for each atom is obtained by a matrix inverse.
In previous implementations, since only the first block of the inverted matrix is needed, the matrix inverse is achieved by the block LU algorithm with the complexity scales with $N_{\mathrm{LIZ}}^2$ \cite{PhysRevLett.75.2867,EISENBACH20172}. The total complexity of LSMS scales with $NN_{\mathrm{LIZ}}^2$, which is linear with respect to total atom number $N$ in the unit cell.

\section{Implementation and Optimization}
In the LSMS method, matrix inverse takes a large proportion of the total computation time.
For example, it takes $92\%$ of the total wall-clock time when the rank of the KKR matrix is 4350. However the matrix inverse can be significantly accelerated by GPUs.

\subsection{System Architecture}
%\textcolor{red}{(I updated info on two B2 GPU and BIL A100 -- I am not sure if you tested on 4 node types or 3.  I can delete the 32GB V100 if you are not including that I am not sure of peak performance using tensor cores for PSC nodes -- should I find that out?)}}
Our GPU benchmarks were performed on four kinds of computing systems: 1) the V100-16G GPU node on Bridges2 at PSC\cite{10.1145/3437359.3465593}; 2) the V100-32GB GPU node on Bridges-2 at PSC; 3) the A100-80G GPU node on the Brain Image Library cluster at PSC\cite{10.1145/3311790.3396653}; 4) the A100-40G GPU node on Lonestar6 at TACC\cite{TACC_LS6}.

Bridges-2 at PSC hosts several models of GPU nodes including 24 GPU nodes with each node having 2x Intel Xeon Gold 6248 20-Core Processor ("Cascade Lake"), with 40 cores on 2 sockets and 16 x 32GB totalling 512GB DDR4-2933 RAM.  There are 8 NVIDIA V100 GPU configured on each node.  Each V100 GPU has 32GB HBM2 memory.

Bridges-2 at PSC also hosts 9 GPU nodes with each node having 2x Intel Gold 6148 20-core Processors with 40 cores on 2 sockets and 12 x 16GB totalling 192GB DDR4-2666 RAM. There are 8 NVIDIA V100 GPU configured on each node.  Each V100 GPU has 16GB HBM2 memory.

The Brain Image Library cluster at PSC has an 8 A100 SXM4 node with 2x EPYC 7543 32-Core Processor ("Milan"), with 64 cores on 2 sockets and 32 x 64GB totalling 2TB RAM DDR4 memory.  There are 8 NVIDIA A100 GPU.  Each A100 GPU has 80GB HBM2 memory.

 Lonestar6 at TACC hosts 32 GPU nodes. Each node has 2x AMD EPYC 7763 64-Core Processor ("Milan"), with 128 cores on two sockets, totally 256GB DDR4-3200 RAM. There are 3 NVIDIA A100 GPUs configured on each node, GPU0 on socket 0 and GPU1,2 on socket 1. Each A100 GPU has 40GB HBM2 memory.%, and has a peak performance of 9.7 TFlops in double precision and 312 TFlops in FP16 precision using the Tensor Cores.   %\textcolor{red}{Added by hliu about lonestar6 GPU:} \textcolor{red}{This section needs input from Ed and Hang ... for the description of the GPU nodes architecture and the system environment of Bridges2, BIL, and Lonestar.  Ed here, I tried to follow the Lonestar template -- let me know if you want me to flush out further}

\subsection{Fortran-C Interface}
\begin{figure}[h]
  \centering
  \includegraphics[width=0.7\linewidth]{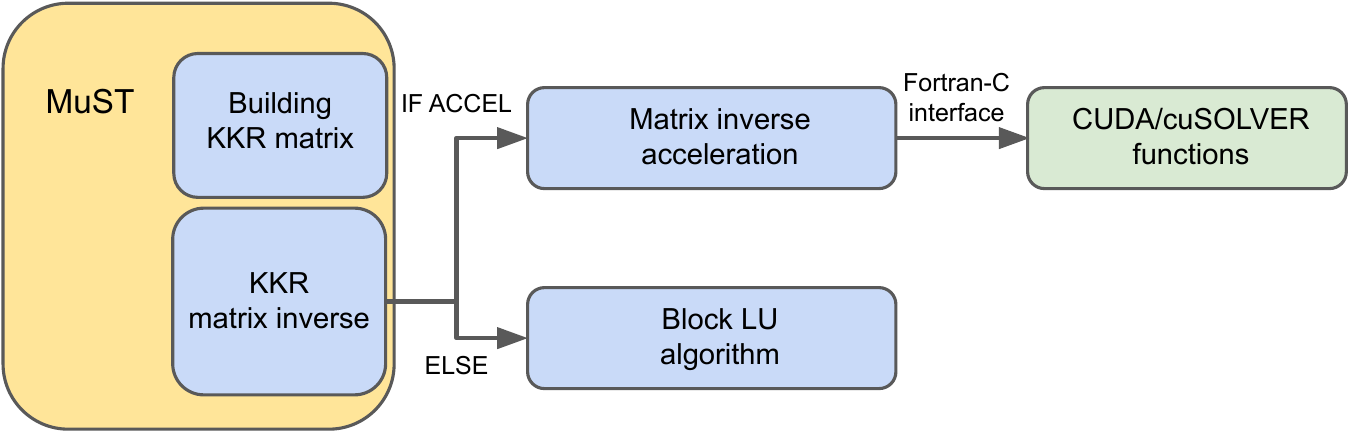}
  \caption{Illustration the procedure of calling functions for matrix inverse in the MuST program. Blue rectangles depict the Fortran subroutines. The green rectangle depicts the C program which calls the CUDA and \texttt{cuSOLVER} functions. }
  \label{procedure}
\end{figure}
The main part of the MuST software is written in Fortran programming language, and the GPU calculation is written in CUDA programming language. To offload the matrix inverse on GPUs, we built a C interface to the CUDA code which allows to exchange data with the main Fortran code.

Fig.(\ref{procedure}) illustrates the procedure of calling functions for matrix inverse in the MuST program. The KKR matrix $M$ is built on CPU, after $M$ is built. The compiling flag \texttt{ACCEL} %$\mathrm{ACCEL}$
is used to determine where the matrix inverse is computed. When \texttt{ACCEL} is undefined or set to \texttt{FALSE} %$\mathrm{ACCEL}=\mathrm{False}$,
the matrix inverse is computed on CPU using the block LU algorithm, while for \texttt{ACCEL} defined to be \texttt{CUDA}. %$\mathrm{ACCEL}=\mathrm{True}$,
A Fortran subroutine initializes the variables, then transferring variables to the routine in CUDA through the Fortran-C interface.

In general, there are four steps in the GPU program: 1) Initialize memory space on GPU; 2) Copy data from the host memory to the GPU memory; 3) Compute the matrix inverse; and 4) Copy the results from the GPU memory back to the host memory. The inverse of the matrix $\underline{X}$ is done by solving the linear equation system: $\underline{X}' \underline{X}=\underline{I}$, where $\underline{I}$ is the identity matrix. When each element in $\underline{X}$ is a complex number with double precision real part and imaginary part, the matrix inverse can be achieved through two \texttt{cuSOLVER} functions: \texttt{cusolverDnZgetrf} and \texttt{cusolverDnZgetrs}, where \texttt{cusolverDnZgetrf} performs the LU factorization of the matrix and \texttt{cusolverDnZgetrs} solves the linear equation.

\section{Performance Evaluations}
We evaluate the performance of the LSMS method implemented in MuST by comparing the wall-clock time with and without GPU accelerations.

\subsection{Test Cases}
There are totally three test cases in our benchmarks: the CoCrFeMnNi high entropy alloy with spin-canting and the NiAu binary random alloy with and without spin-canting.
The concentraion of each element in CoCrFeMnNi is the same as $20\%$, and the concentration of Ni and Au in the NiAu is $30\%$ and $70\%$ respectively.
The total number of atoms is 56 for CoCrFeMnNi and 64 for NiAu.

In our test cases $l_{max}=4$,  $N_{\mathrm{LIZ}}=135$ for CoCrFeMnNi and $N_{\mathrm{LIZ}}=249$ for NiAu.
The rank of the KKR matrix for CoCrFeMnNi alloy is 6750. The rank of the KKR matrix for NiAu alloy without and with spin-canting is 6225 and 12450, respectively.
%The KKR matrix size for each test case is depicted in Table.(\ref{LSMS_KKR_matrix_size}).
Each element in the KKR matrix is a complex number with double precision real part and imaginary part.
%\begin{table}[]
%\begin{tabular}{|c|c|c|c|}
%\hline
%Test Case       & CoCrFeMnNi(spin-canted) & NiAu(non spin-canted) & NiAu(spin-canted) \\ %\hline
%KKR Matrix Size & $6750\times 6750$ & $6225\times 6225$ & $12450\times 12450$ \\ \hline
%\end{tabular}
%\caption{The KKR matrix size in the LSMS calculations for our three test cases. }
%\label{LSMS_KKR_matrix_size}
%\end{table}

\subsection{Performance Comparisons}
\begin{figure}[h]
  \centering
  \includegraphics[width=0.8\linewidth]{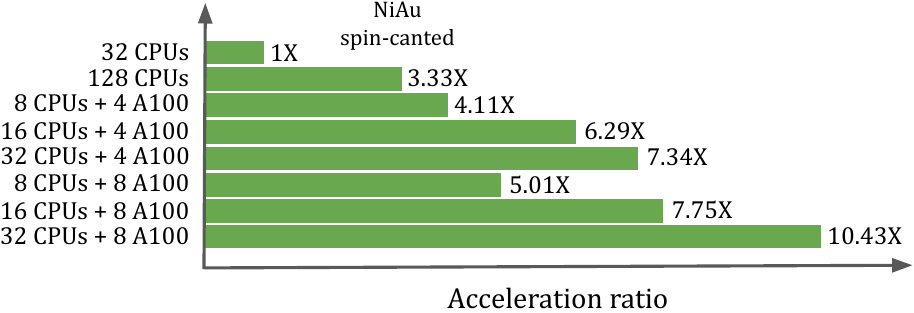}
  \caption{The performance comparisons for various configurations of CPUs and GPUs. The CPU number is the total number of the MPI ranks. The MPI ranks are evenly distributed on multiple GPUs. }
  \label{performance_comparisons}
\end{figure}
Here we present the performance results on various computing systems within different configurations of GPUs and CPUs. We set the OpenMP thread number to 1 through our benchmarks.
The detailed data of our benchmarks is depicted on Table.(\ref{Performance_results}).
The number in the bracket denotes the ratio of the wall-clock time for the GPU data transfer divides by the GPU compute.

Fig.(\ref{performance_comparisons}) depicts the comparisons of wall-clock time of the MuST program excluding the I/O operations. The reference CPU benchmarks were performed on the Regular Memory CPU Nodes on Bridges2 at PSC, and the CPU type is the AMD EPYC 7742. The MuST program was compiled with the AMD optimized math library AOCL for the reference CPU benchmarks.

In the figure, the baseline is the performance of 32 CPU cores of the CPU node. Apparently, GPUs are able to accelerate the calculation with a significant amount of speedup ratio.
For the case of using either 4 or 8 GPUs, the job running on 16 CPU cores shows roughly 1.5 times faster than running on 8 CPU cores. Running on 32 CPU cores are roughly 1.2 times and 1.3 times faster than running on 16 CPU cores for the case of using 4 and 8 GPUs, respectively. For the case of running on either 8 or 16 CPU cores, using 8 GPUs are roughly 1.2 times faster than using 4 GPUs. However when running 32 CPU cores, using 8 GPUs is roughly 1.4 times faster than using 4 GPUs.

\section{Discussion and Conclusion}
We demonstrated that the Green's function based electronic structure software MuST can achieve high acceleration ratio on acceleration cards like NVIDIA GPUs. Our work opens the door to simulate a large disordered alloy system in a moderate time. We would like to point out that the target for GPU offloading is also a good accelerating target on multi-core CPUs. Our test results using threaded BLAS libraries from Intel MKL and NVHPC shows sub linear speedup in terms of thread numbers on Intel and AMD multi-core CPUs. This helps to achieve same scaling performance with much less number of MPI tasks, and more efficient parallel I/O in future.

In this work the GPU acceleration is only demonstrated on the LSMS method, which takes a cluster approximation to achieve the linear scaling with respect to the number of atoms in the unit cell. In fact, it is quite straightforward to offload the KKR matrix inverse on GPUs to accelerate the original KKR calculation. Further code acceleration is expected if the KKR matrix is constructed on GPUs, instead of on CPUs, in which case the time for data transfer from CPU to GPU will be significantly reduced.
Furthermore, the matrix inverse acceleration is not limited to NVIDIA GPUs, the acceleration cards from other vendors are promising to deliver competitive acceleration results. Investigating the performance on other acceleration cards is one of our future goals.

%%
%% The acknowledgments section is defined using the "acks" environment
%% (and NOT an unnumbered section). This ensures the proper
%% identification of the section in the article metadata, and the
%% consistent spelling of the heading.
\begin{acks}
X.L. thanks usefull discussions with Markus Eisenbach, Vishnu Raghuraman and Michael Widom. E.H. thanks Tod Pike at PSC for setting up access to the Brain Image Library A100 GPU node. H.L. thanks Dr. Junjie Li for providing his tool to profile BLAS calls in the code and tips of MPI task to GPU binding.  This work was supported by the National Science Foundation through the OAC-2139536 Characteristic Science Applications for the Leadership Class Computing Facility award. The MuST package is the product of an open source project supported in part by NSF Office of Advanced Cyberinfrastructure and the Division of Materials Research within the NSF Directorate of Mathematical and Physical Sciences under award number 1931367, 1931445, and 193152.
\end{acks}

%%
%% The next two lines define the bibliography style to be used, and
%% the bibliography file.
\bibliographystyle{ACM-Reference-Format}
\bibliography{references}

%%
%% If your work has an appendix, this is the place to put it.
\appendix

\begin{landscape}
\begin{table}[]
\begin{tabular}{|c|cccc|cc|}
\hline
\multirow{5}{*}{\begin{tabular}[c]{@{}c@{}}CoCrFeMnNi\\ spin-canted\\ (on TACC-LS6 2-A100-40G)\end{tabular}}      & GPU number/MPI rank number & 8            & 16          &             & \begin{tabular}[c]{@{}c@{}}MPI rank number\\ (=CPU core number)\end{tabular} &       \\
                                                                                                                  & 1                          & 3256(0.097)  &             &             & 16                                                                           & 7966  \\
                                                                                                                  & 2                          & 2283(0.106)  & 1401(0.074) &             & 28                                                                           & 4647  \\
                                                                                                                  &                            &              &             &             & 56                                                                           & 2597  \\
                                                                                                                  &                            &              &             &             & 112                                                                          & 1785  \\ \hline
\multirow{4}{*}{\begin{tabular}[c]{@{}c@{}}NiAu\\ non spin-canted\\ (on TACC-LS6 2-A100-40G)\end{tabular}}        & GPU number/MPI rank number & 8            & 16          &             & \begin{tabular}[c]{@{}c@{}}MPI rank number\\ (=CPU core number)\end{tabular} &       \\
                                                                                                                  & 1                          & 1948(0.089)  &             &             & 16                                                                           & 4680  \\
                                                                                                                  & 2                          & 1347(0.0815) & 1014(0.077) &             & 32                                                                           & 2489  \\
                                                                                                                  &                            &              &             &             & 64                                                                           & 1447  \\ \cline{1-5}
\multirow{5}{*}{\begin{tabular}[c]{@{}c@{}}NiAu\\ non spin-canted\\ (on PSC-Bridges2 8-V100-16G )\end{tabular}}   & GPU number/MPI rank number & 8            & 16          & 32          & 128                                                                          & 999   \\
                                                                                                                  & 1                          &              &             &             &                                                                              &       \\
                                                                                                                  & 2                          & 1761(0.137)  &             &             &                                                                              &       \\
                                                                                                                  & 4                          & 1279(0.178)  & 979(0.158)  &             &                                                                              &       \\
                                                                                                                  & 8                          & 936(0.305)   & 702(0.181)  & 593(0.180)  &                                                                              &       \\ \hline
\multirow{4}{*}{\begin{tabular}[c]{@{}c@{}}NiAu\\ spin-canted\\ (on PSC-Bridges2 8-V100-32G)\end{tabular}} & GPU number/MPI rank number & 8            & 16          &             & \begin{tabular}[c]{@{}c@{}}MPI rank number\\ (=CPU core number)\end{tabular} &       \\
                                                                                                                  & 2                          &              &             &             & 32                                                                           & 19002 \\
                                                                                                                  & 4                          & 6271(0.113)  &             &             & 64                                                                           & 11069 \\
                                                                                                                  & 8                          & 3924(0.192)  & 3375(0.130) &             & 128                                                                          & 5695  \\ \cline{1-5}
\multirow{4}{*}{\begin{tabular}[c]{@{}c@{}}NiAu\\ spin-canted\\ (on PSC-BIL 8-A100-80G )\end{tabular}}            & GPU number/MPI rank number & 8            & 16          & 32          &                                                                              &       \\
                                                                                                                  & 2                          & 5528(0.101)  & 4543(0.108) &             &                                                                              &       \\
                                                                                                                  & 4                          & 4618(0.104)  & 3022(0.117) & 2589(0.134) &                                                                              &       \\
                                                                                                                  & 8                          & 3790(0.194)  & 2452(0.129) & 1821(0.148) &                                                                              &       \\ \hline
\end{tabular}
\caption{The performance comparisons on various systems with different configurations of GPUs and CPUs. The value is the wall-clock time for running the MuST program except IO operations, the unit is seconds. The ratio in the bracket is the wall-clock time for GPU memory transfer divides by the wall-clock time for GPU compute. }
\label{Performance_results}
\end{table}
\end{landscape}

\end{document}